\documentclass[a4paper,11pt]{article}
\pdfoutput=1 % if your are submitting a pdflatex (i.e. if you have
\usepackage{jinstpub} % for details on the use of the package, please
\title{Improving Reflexive Surfaces Efficiency with Genetic Algorithms}

\author[1]{A. Steklain\note{Corresponding author.},}
\author{M. Adames and}
\author{F. Ganacim}

\affiliation{Department of Mathematics, Universidade Tecnol\'{o}gica Federal do Paran\'{a}\\Av. Sete de Setembro, 3165, Brazil}

\emailAdd{steklain@utfpr.edu.br}

\abstract{We propose using a Genetic Algorithm to improve the efficiency of reflexive surfaces in devices where the receiver's position is different from the classic parabolic antenna. With this technique, we show that we can improve the efficiency of the ARAPUCA photodetector.}

\keywords{Simulation methods and programs; Photon detectors for UV, visible and IR photons (vacuum) (photomultipliers, HPDs, others); Cryogenic detectors}

\begin{document}
\maketitle
\flushbottom

\section{Introduction}
\label{sec:intro}

\begin{figure}[!ht]
    \centering
    \includegraphics[width=0.4\textwidth]{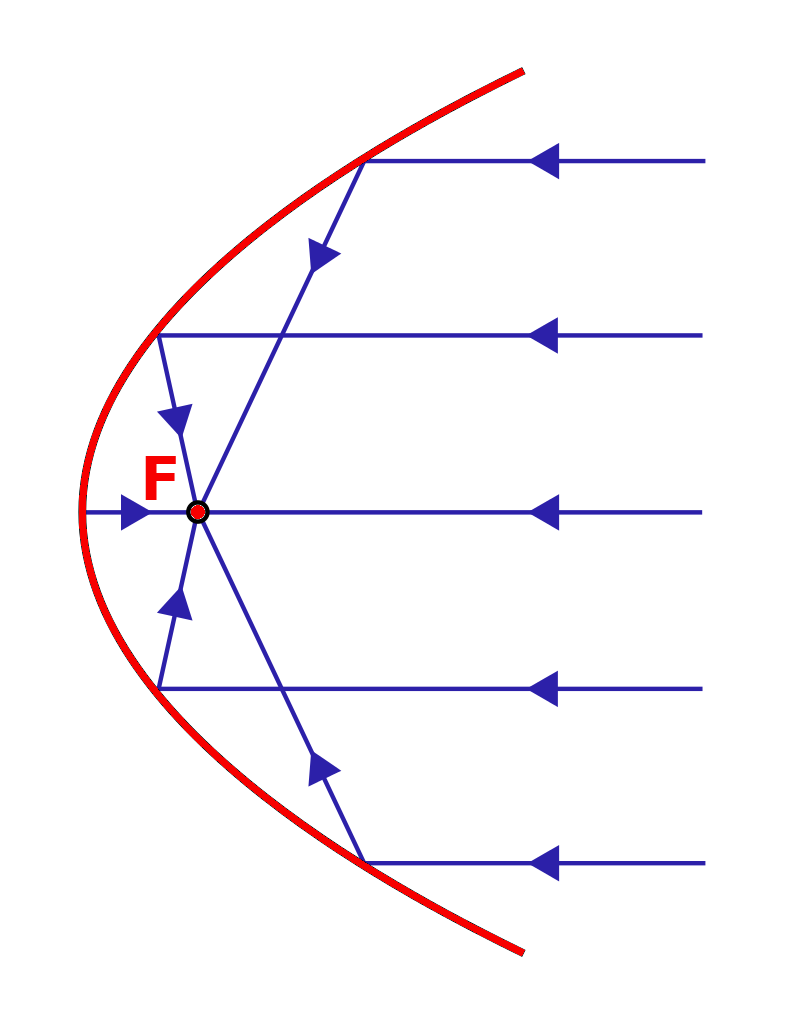}
    \caption{Reflection in a surface shaped as a paraboloid. Parallel rays are reflected towards the focus.}
    \label{fig:paraboloid}
\end{figure}

In general, the construction of telescopes and antennas uses reflective surfaces as part of the design. For such devices, traditionally, the surface is shaped like a paraboloid, so a parallel electromagnetic beam can be reflected by the paraboloid surface towards the focal point, maximizing efficiency. We show this guiding principle in Figure~\ref{fig:paraboloid}. It is possible to find the same concept in acoustics, with metamaterials designed to induce parabolic phase gradients \cite{Song2016}. 

New devices using reflective surfaces have been proposed in the last few years. In particular, a new generation of photodetectors like ARAPUCA and X-ARAPUCA use reflections in a photon trap to improve the efficiency \cite{Machado2016, Machado2018}. Such devices, in general, have several constraints in their construction, so even an off-axis paraboloid solution can not be implemented. On the other hand, photons can reflect in the internal walls more than once before they get detected. In such cases, a new surface that maximizes the number of reflections directed to a given area, such as a photomultiplier, is desirable. Although there are no analytical methods to obtain such a surface, it is possible to use numerical optimization algorithms to achieve this goal.

Metaheuristic algorithms are used mainly for solving optimization problems \cite{AbdelBasset2018}. Among these methods, we find metaphor-based algorithms inspired by biological or physical phenomena \cite{Katoch2021}. We can also classify these algorithms between those using a single candidate and those using a population of candidates. Single-solution-based algorithms, such as simulated annealing, may be stuck into local optima. On the other hand, algorithms using populations can maintain diversity and avoid such problems. The main disadvantage of population-based algorithms is their computational cost, so using as few iterations as possible is necessary. One of the best-known metaheuristic algorithms based on populations is the Genetic Algorithm (GA).

The idea behind GA is the biological process known as evolution, combining natural selection and genetics \cite{goldberg1989,kubat2017}. This algorithm mimics the Darwinian theory, where the fittest individuals have a more significant probability of surviving and reproducing. Contrary to natural selection in Nature, individuals do not need strength or sharp teeth to be selected. We can define the fitness function using any quantity we want to maximize or minimize, including, for instance, efficiency in photon detection.

In this work, we will use GA to find a reflective surface that maximizes the efficiency of an ARAPUCA device. We model the genetic code using the coefficients obtained from the discrete Fourier transform of the surface. By combining these coefficients, we can find optimized surfaces improving the efficiency of the photodetector. We organize this work as the following. In Section~\ref{sec:arapuca}, we present the concept of the ARAPUCA device and describe our Python-based model. In Section~\ref{sec:ga}, we present the main routines we use in the GA algorithm. In Section~\ref{sec:results}, we discuss our results. Finally, in Section~\ref{sec:conclusion}, we present our main conclusions.

%\cite{Sang2021}

\section{ARAPUCA Simulation}
\label{sec:arapuca}

The main advance provided by ARAPUCA technology \cite{Machado2016} is the increasing of the photodetection setup active coverage without merely using more active photosensors. The ARAPUCA system combines a passive light collector and active devices, in which the photon remains trapped inside the apparatus until its detection.

\begin{figure}
    \centering
    \includegraphics[width=0.7\textwidth]{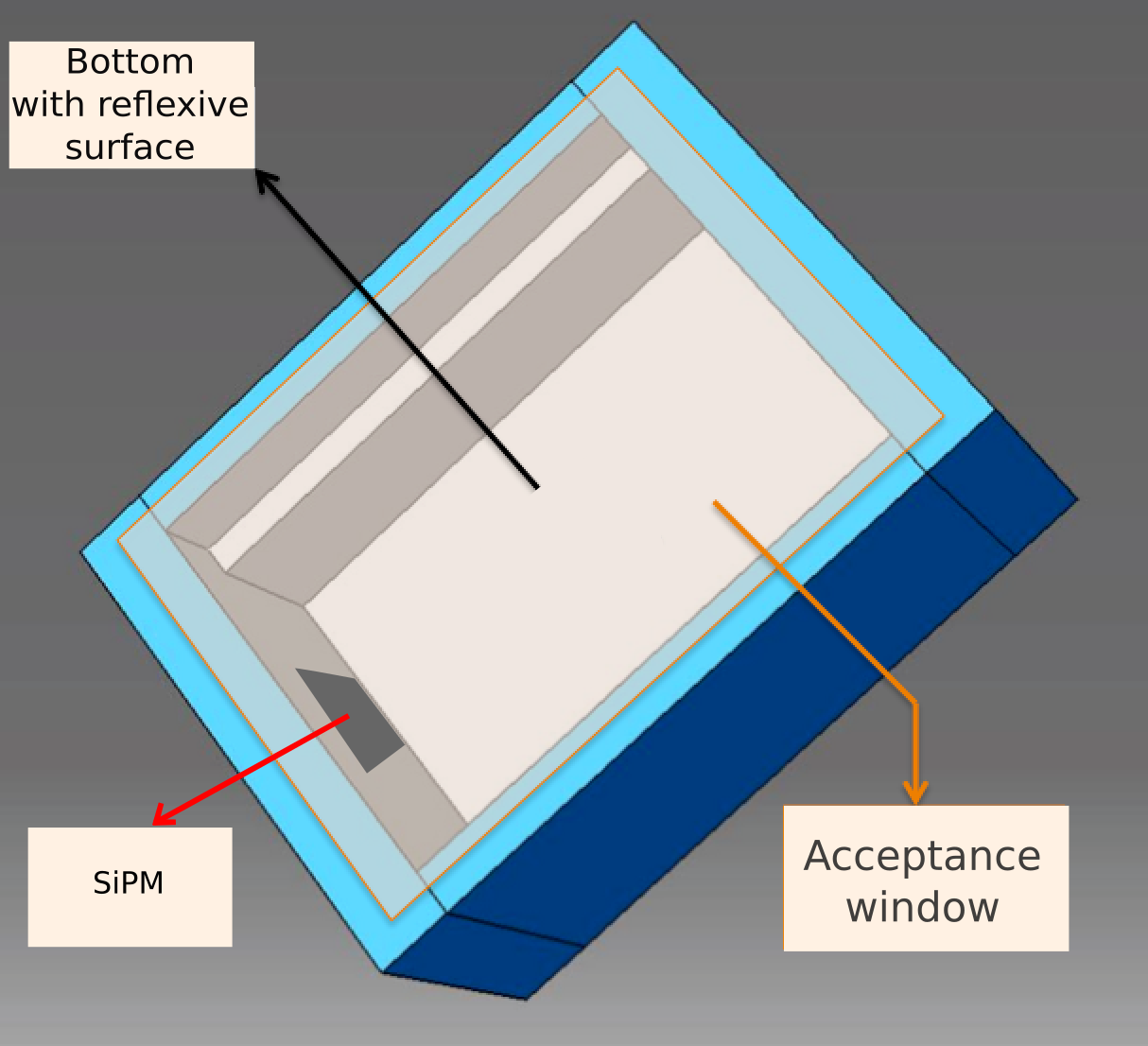}
    \caption{design of an ARAPUCA device. The incident photon passes through the acceptance window and is converted to a new wavelength. As the window is reflexive to this wavelength, it remains trapped inside the box until it gets detected by the active detector (SiPM) located in one of the laterals. Adapted from \cite{Machado2016}.}
    \label{fig:arapuca}
\end{figure}

We present a design of the ARAPUCA system in Figure~\ref{fig:arapuca}. A simplified working scheme is the following. The acceptance window comprises a dichroic filter and a wavelength shifter (WLS). Photons with a determined range of wavelengths are transparent to the dichroic filter and pass through the window. These photons are absorbed by the WLS and re-emitted with a different wavelength. The dichroic filter is perfectly reflexive to the new photon trapped inside the box. The window and the inner walls reflect the photon until it gets detected by the photosensor (SiPM) located at one of the walls.

There are several modifications proposed to increase the ARAPUCA efficiency. The most successful version is known as X-ARAPUCA \cite{Machado2018}. In this version, the wavelength shift occurs inside a light guide introduced inside the ARAPUCA box. The re-emitted photons propagate in all directions, and total internal reflections guide the photons to the photosensor. Other geometrical designs have been studied to increase the efficiency of this system, but all of them use a reflexive plane wall at the bottom of the device. It is impossible to modify the geometry of the acceptance window because of how its produced, but the bottom reflexive surface can be modeled in mechanical ways or constructed by 3D printing. Optimizing the reflexive bottom surface of the original ARAPUCA will be the goal of this study.

In this work, we simulate ARAPUCA-like devices using Python combined with a C\texttt{++} ray-tracing routine, as this approach proves to be simple and fast. Nevertheless, a complete simulation including all optical properties using Geant4 is desirable to confirm these results.

\begin{figure}
    \centering
    \includegraphics[width=0.8\textwidth]{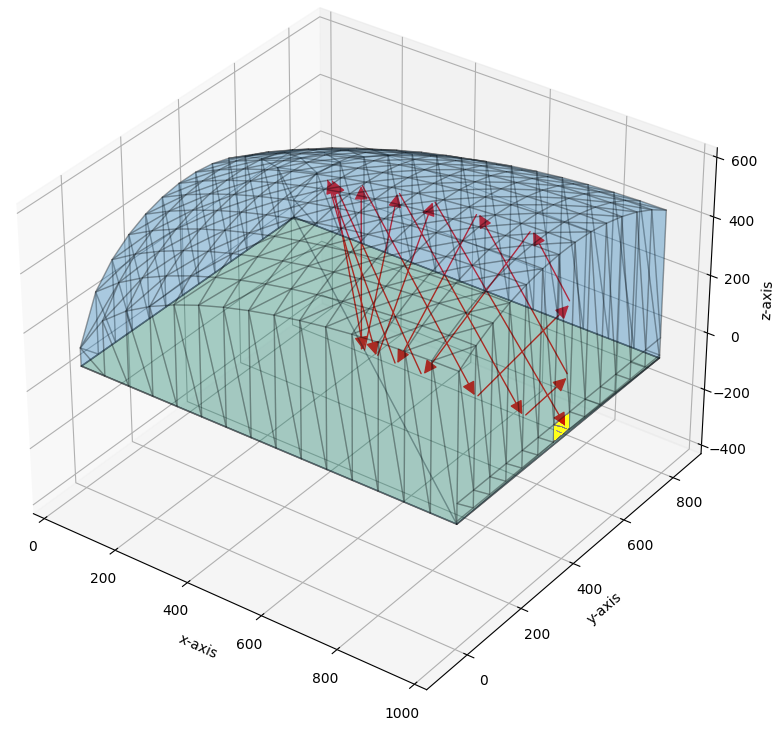}
    \caption{ARAPUCA simulation in Python. The photons, represented by red arrows, are emitted from the acceptance window (green) below and reflected by the walls (blue). The photosensor is the small yellow square located at the wall.}
    \label{fig:arapuca_sim}
\end{figure}

The simulation consists of a box composed below by a plane representing the acceptance window and at the top by a reflexive mesh surface. In this mesh, the nodes are given by $(x,y,z)$ points where the coordinates $(x,y)$ are fixed, and $z(x,y)$ is allowed to change so that the surface can be interpreted as the plot of a two-dimensional function. The two surfaces are connected by four walls that close the box, as shown in Figure~\ref{fig:arapuca_sim}. The photosensor is a small square that will be selected in the box to count and stop the photons that hit that region. The photons are generated in the acceptance window and propagate only inside the box, being reflected by all surface points until they reach the photosensor or a maximum of 50 reflections. We only count the photons that reach the photosensor. 

\section{GA Design for Surface Optimization}
\label{sec:ga}

\begin{figure}
    \centering
    \includegraphics[width=0.6\textwidth]{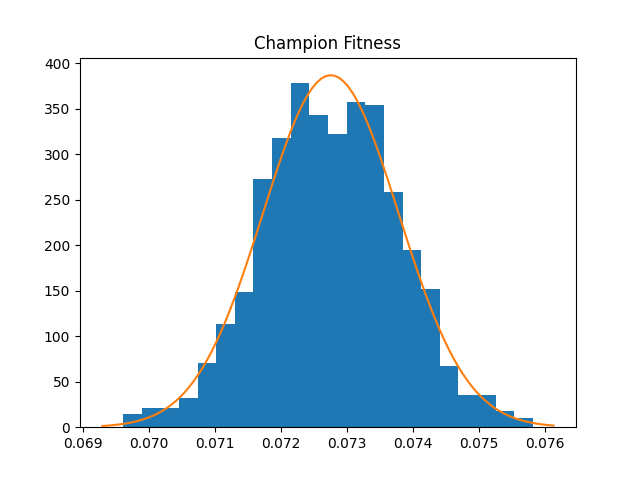}
    \caption{Histogram generated from different efficiency measures of the same box.}
    \label{fig:histoefficiency}
\end{figure}

The main goal of GA is to find an extreme value for the fitness function. In this work, the fitness is the photodetector efficiency, defined as the ratio between the number of detected and generated photons. The generated number of photons is fixed, but the number of photons detected is a random variable, as the trajectories of the photons are randomly generated. It implies that the same box can have different values for efficiency. In Figure~\ref{fig:histoefficiency}, we show a histogram from different efficiency measures for the same box. If the fitness of two individuals is close, there is uncertainty about which one is the largest. To address this problem, we use a statistical test to decide which one is larger. To make this decision, we use a nonparametric test, namely the sign test \cite{Zhai1996}. This test has the form
\begin{equation*}
    T = \frac{1}{M} \sum _{\ell=1} ^M sgn \left( Z^{(\ell)} \right),
\end{equation*}
where $sgn(x)=1$ if $x>0$ and $sgn(x)=-11$ if $x<0$. $M$ is the number of samples of the variable $Z$ corresponding to the difference between the fitness of two separate boxes. 

GA comprises several different modules that must be chosen and adapted to each problem. For some systems, this adaptation is straightforward, but for others, the choices are far from trivial. The main components of GA are the Fitness function, Chromosomes, Initial Population, Survival Game, Mating Operator, Recombination, and Mutation. 

The Chromosomes are the most sensitive component of the algorithm, as they must be balanced between the freedom to generate individuals with distinct characteristics and the convergence of the method. Although the obvious choice seems to be the $z$ coordinate of the surface points, such a method did not provide good results. The main reason is the abrupt changes in the mesh and the reflection angles. Taking advantage of the fact that the surface can be written as a function of two variables, we consider the Fourier decomposition of such a surface instead. The Chromosomes are the components of the two-dimensional Fast Fourier Transform applied in the $z$ values of the mesh.

The Initial Population also must be chosen to guarantee the necessary genetic diversity but can be designed to outperform randomly generated individuals \cite{kubat2017}. Instead of using random individuals, we use a family of spherical sections. This choice is justified because, in terms of Fourier decomposition, these populations still have genetic variability, and on the other hand, the smoothness speed up the algorithm convergence.

The Survival Game is responsible for selecting which individuals will be preselected to survive and have a chance to find a partner. Although several algorithms use some higher level of randomization in a ``wheel of fortune,'' we select a percentage of the population based on the fitness function value.

For the Mating Operator, we use the strategy shown in \cite{kubat2017}. Individuals with larger fitness choose their partners randomly, but the probability of each individual being chosen as a partner is proportional to their fitness. It guarantees that individuals with small fitness can be chosen as partners, although with a smaller chance.

The Recombination used is obtained from \cite{Toledo2014}. It improves the simple crossover of parents
\begin{align*}
    parent1 &= [m_1,m_2,m_3,..., m_N]\\
    parent2 &= [d_1,d_2,d_3,..., d_N]
\end{align*}
where the crossover point $\alpha$ is a random integer from 1 to $N$. The offspring generated is then
\begin{align*}
    offspring1 &= [m_1,m_2,...,m_{\alpha-1},g_1,d_{\alpha+1}...,d_N]\\
    offspring2 &= [d_1,d_2,...,d_{\alpha-1},g_2,m_{\alpha+1}...,m_N]
\end{align*}
where in the crossover point, we perform a linear combination between the values of the chromosome in position $\alpha$ of the two parents
\begin{align*}
    g_1 = m_\alpha - \beta (m_\alpha-d_\alpha)\\
    g_2 = d_\alpha + \beta (m_\alpha-d_\alpha)
\end{align*}
where $\beta$ is a random number between 0 and 1.

Finally, the Mutation is performed randomly according to a fixed probability. The intensity of the Mutation is determined using another random number with a maximum value defined \emph{a priori}. 

In this work, we use a population consisting of 200 individuals.

\section{Results}
\label{sec:results}

\begin{figure}
    \centering
    \includegraphics[width=0.8\textwidth]{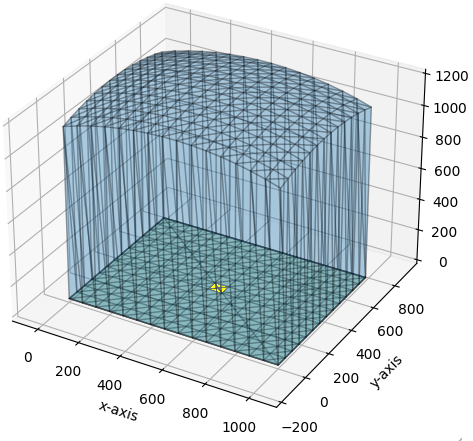}
    \caption{Parabolic antenna simulation in Python. The photons are emitted from the acceptance window (green) and reflected by the walls (blue). The photosensor is the small yellow square located in the center of the acceptance window.}
    \label{fig:parab_antenna}
\end{figure}

As we stated before, for a parallel beam, the reflector antenna (or telescope), the problem has a family of optimal solutions consisting of paraboloids such that the detector is located on the vertex. In this case, the efficiency will be 1, i.e., all the reflected photons will be collected in the detector. Some photons can be lost for a discretized mesh, but the efficiency is still high. We simulate such a setup from the ARAPUCA simulation by positioning the detector in the center of the acceptance window, as shown in Figure~\ref{fig:parab_antenna}.

\begin{figure}
    \centering
    \includegraphics[width=\textwidth]{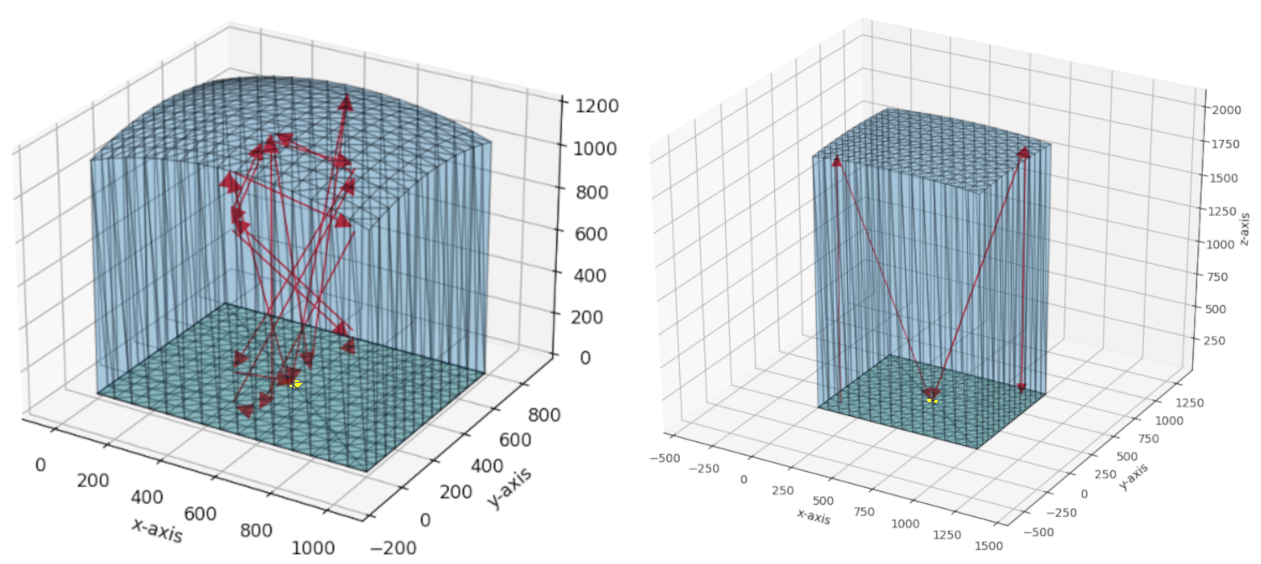}
    \caption{Reflexive surface before and after GA optimization for a parallel light beam. Before the optimization, the surface is a hemisphere, and the efficiency is 0.06. After 250 generations, the surface is a paraboloid with an efficiency of 0.93.}
    \label{fig:resultspa}
\end{figure}

The reflector antenna is a good test for the proposed GA. We start with a family of spherical sections centered in the detector and with different radii as the initial population. For such a family of surfaces, the maximum efficiency is 0.06. With the GA application, the boxes obtained have a maximum efficiency of 0.93 after 250 generations. Sample boxes before and after GA are shown in Figure~\ref{fig:resultspa}.

For ARAPUCA, we use a family of spherical surfaces centered in the detector as the initial population. Nevertheless, as the detector is located at one of the lateral walls of the box, the spherical sections look different from the case of the reflector antenna.

\begin{figure}
    \centering
    \includegraphics[width=\textwidth]{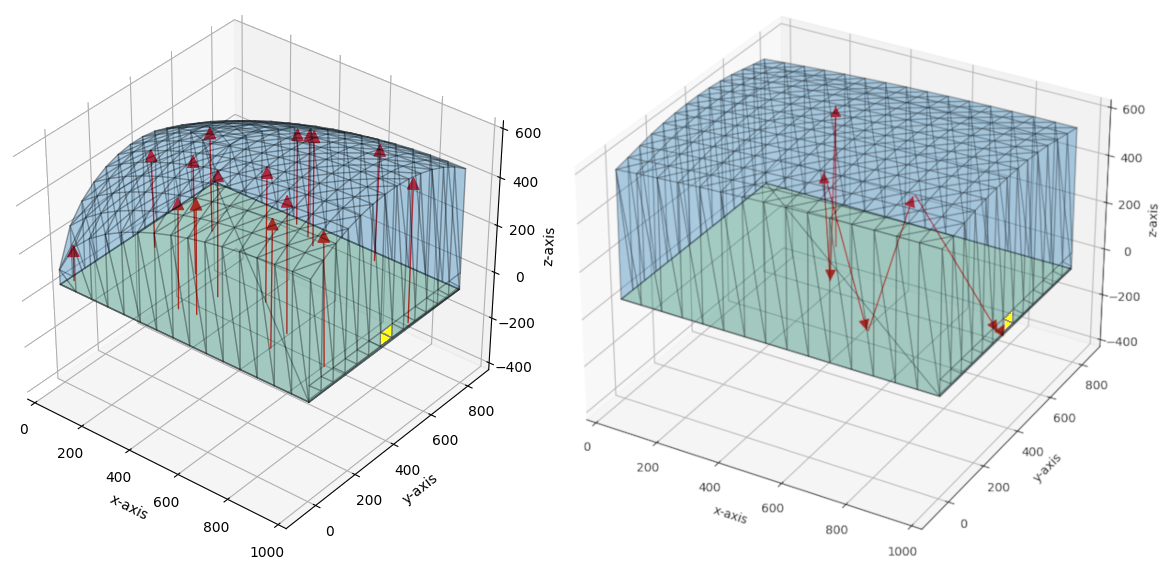} 
    \caption{Reflexive surface before and after GA optimization for a parallel light beam. Before the optimization, the efficiency is 0.14. After 1000 generations, the efficiency is 0.52.}
    \label{fig:resultsarapuca}
\end{figure}

In the first setup, we use a parallel beam of photons emitted perpendicularly from the window at the bottom of the box. In this case, the maximum efficiency of the initial population was 0.14. After applying GA, the maximum efficiency is 0.52 after 1000 generations. Although it was impossible to reach a more significant efficiency, there was a great improvement considering the initial design (hemispherical sections). Sample boxes before and after GA are shown in Figure~\ref{fig:resultsarapuca}.

\begin{figure}
    \centering
    \includegraphics[width=\textwidth]{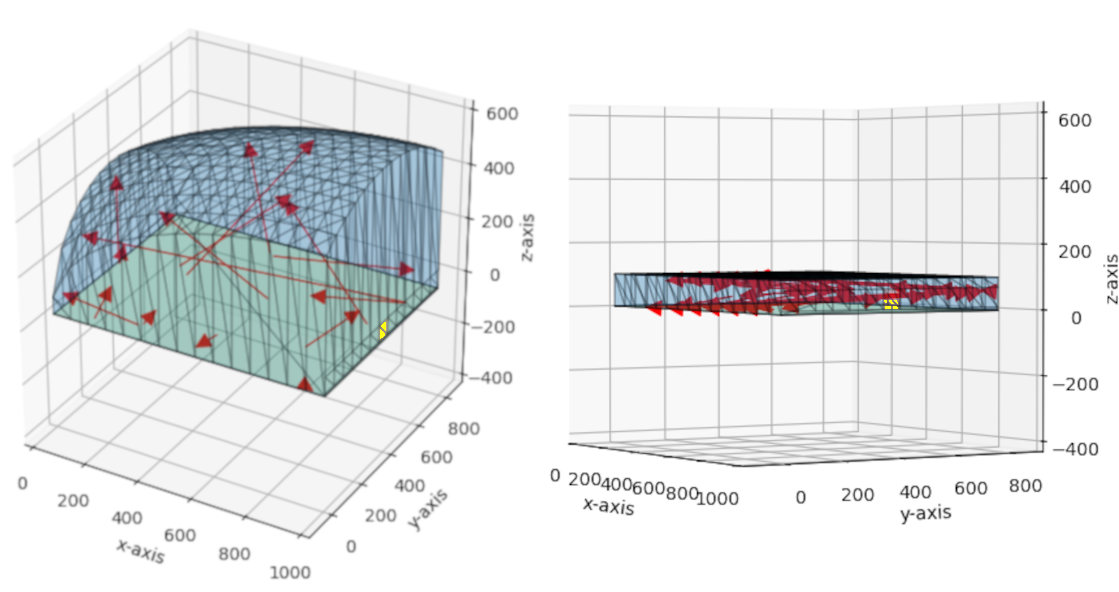} 
    \caption{Reflexive surface before and after GA optimization for a random light beam. Before the optimization, the efficiency is 0.06. After 1000 generations, the efficiency is 0.14. In this case, the GA algorithm can not find an alternative for the reflexive surface, and a plane surface is the best option.}
    \label{fig:resultsarapucarandom}
\end{figure}

In the second setup, we use photons emitted in random directions from the window at the bottom of the box. In this case, the maximum efficiency was initially 0.06, and after the application of GA, we found that the efficiency achieved a maximum value of 0.14. Sample boxes before and after GA are shown in Figure~\ref{fig:resultsarapucarandom}. It is interesting to note that for random directions, the best individual consists of a plane reflective surface, the same design already used for the original ARAPUCA.

\section{Conclusions}
\label{sec:conclusion}

The optimization of reflective surfaces using GA is possible using 2D Fourier components as chromosomes. This method was proven successful in the optimization of a reflective antenna. For the ARAPUCA system, this method improved the reflective surface for a parallel beam of light. When light is generated from random directions, the plane surface already in use for ARAPUCA is the best choice.

For the following steps, we will improve the ARAPUCA simulation to make it more realistic in the physical sense. One possibility is to use Chroma or Geant4 to perform the light propagation. Another step is to consider a GA optimization of  X-ARAPUCA. In this case, we can consider not only the reflective surface but also the height of the light guide.

\acknowledgments

We thank DUNE-BR, especially Ana Machado, Ettore Segreto, and Gustavo Valdiviesso, for all the valuable discussions.


\bibliographystyle{JHEP}
\bibliography{cas-refs}

\providecommand{\href}[2]{#2}\begingroup\raggedright\begin{thebibliography}{1}

\bibitem{Song2016}
K.~Song, J.~Kim, S.~Hur, J.-H.~Kwak, S.-H.~Lee and T.~Kim, \emph{{Directional
  Reflective Surface Formed via Gradient-Impeding Acoustic Meta-Surfaces}},
  \href{https://doi.org/10.1038/srep32300}{\emph{Scientific Reports} {\bfseries
  6} (2016) 32300}.

\bibitem{Machado2016}
A.A.~Machado and E.~Segreto, \emph{{ARAPUCA a new device for liquid argon
  scintillation light detection}},
  \href{https://doi.org/10.1088/1748-0221/11/02/C02004}{\emph{Journal of
  Instrumentation} {\bfseries 11} (2016) }.

\bibitem{Machado2018}
A.~Machado, E.~Segreto, D.~Warner, A.~Fauth, B.~Gelli, R.~M{\'{a}}ximo et~al.,
  \emph{{The X-ARAPUCA: an improvement of the ARAPUCA device}},
  \href{https://doi.org/10.1088/1748-0221/13/04/C04026}{\emph{Journal of
  Instrumentation} {\bfseries 13} (2018) C04026}
  [\href{https://arxiv.org/abs/arXiv:1804.01407v1}{{\ttfamily
  arXiv:1804.01407v1}}].

\bibitem{AbdelBasset2018}
M.~Abdel-Basset, L.~Abdel-Fatah and A.K.~Sangaiah, \emph{Chapter 10 -
  metaheuristic algorithms: A comprehensive review},  in \emph{Computational
  Intelligence for Multimedia Big Data on the Cloud with Engineering
  Applications}, A.K.~Sangaiah, M.~Sheng and Z.~Zhang, eds., Intelligent
  Data-Centric Systems, pp.~185--231, Academic Press (2018),
  \href{https://doi.org/https://doi.org/10.1016/B978-0-12-813314-9.00010-4}{DOI}.

\bibitem{Katoch2021}
S.~Katoch, S.S.~Chauhan and V.~Kumar, \emph{{A review on genetic algorithm:
  past, present, and future}}, Multimedia Tools and Applications (2021),
  \href{https://doi.org/10.1007/s11042-020-10139-6}{10.1007/s11042-020-10139-6}.

\bibitem{goldberg1989}
D.E.~Goldberg, \emph{Genetic Algorithms in Search, Optimization, and Machine
  Learning}, Addison-Wesley, New York (1989).

\bibitem{kubat2017}
M.~Kubat, \emph{An Introduction to Machine Learning}, Springer International
  Publishing, Chan (2017).

\bibitem{Zhai1996}
W.~Zhai, P.~Kelly and W.B.~Gong, \emph{{Genetic algorithms with noisy
  fitness}},
  \href{https://doi.org/10.1016/0895-7177(96)00068-4}{\emph{Mathematical and
  Computer Modelling} {\bfseries 23} (1996) 131}.

\bibitem{Toledo2014}
C.F.~Toledo, L.~Oliveira and P.M.~Fran{\c{c}}a, \emph{{Global optimization
  using a genetic algorithm with hierarchically structured population}},
  \href{https://doi.org/10.1016/j.cam.2013.11.008}{\emph{Journal of
  Computational and Applied Mathematics} {\bfseries 261} (2014) 341}.

\end{thebibliography}\endgroup
\end{document}